\documentclass[%
 reprint,
 amsmath,amssymb,
 aps,
]{revtex4-1}

\usepackage{graphicx}
\usepackage{dcolumn}
\usepackage{bm}

\begin{document}

\preprint{APS/123-QED}

\title{Critical Parameterisation of the Brill Wave Initial Value Problem}

\author{Andrew Masterson}
\affiliation{%
Department of Physics and Astronomy \\ University of Calgary (\href{mailto:masterson.andrew@gmail.com}{masterson.andrew@gmail.com})
}%


\author{David Hobill}
\affiliation{
Department of Physics and Astronomy \\ University of Calgary (\href{mailto:hobill@ucalgary.ca}{hobill@ucalgary.ca}) 
}%


\date{\today}

\begin{abstract}
A numerical framework to explore the Brill Wave Initial Value Problem is presented along with a review of some of the theoretical predictions concerning Brill waves.  It is demonstrated that there is an agreement between the numerically observed phenomena and theory, namely that the IVP parameterisation of the metric function $q$ has a minimum and maximum amplitude for which a solution to the Hamiltonian for the metric function $\phi$ exists, and a critical regime in the middle for which there are and are not apparent horizons present.
\end{abstract}

\maketitle


\section{Introduction}
In 1959, Brill \cite{Brill} was able to prove a positive energy theorem in general relativity (GR) for a a particular vacuum solution to the initial value problem (IVP). This demonstrated that general relativity, unlike Newtonian gravitational theory, could possess non-trivial, regular gravitational fields in the absence of massive material sources.  That such a counterexample to Mach's principle exists in relativistic gravity is due to the
existence of a non-vanishing Weyl tensor whose components are determined by the dynamical gravitational degrees of freedom in the vacuum. 

In order to explore the dynamics of Brill wave evolution, a numerical code was developed \cite{MastersonPHD} that allows one to specify a wide range of initial values and the explore subsequent evolution of the vacuum gravitational field. The code was thoroughly tested for accuracy, convergence and stability.  The results of the initial value problem (IVP) solutions obtained from this code were analyzed and compared to the theoretical predictions of {\'O} Murchadha \cite{murchada-1993dgr2.conf..210O}. This paper is organized as follows: the next section briefly reviews the 3+1 splitting of spacetime and in particular how it relates to the problem posed by Brill.  Section III discusses the Brill initial value problem and different functional forms of the free data that satisfy the conditions that an asymptotically vacuum solution must satisfy. The section that follows introduces the method used to solve the IVP for the different cases presented. The solutions to the IVP can, for certain choices of the parameters that appear in the free data contain apparent horizons and Section V presents the
method used to find the trapped surfaces and apparent horizons that exist on the initial time slice. A comparison of the numerical results with some theoretical predictions made by 
{\'O} Murchadha are made in Section VI and based upon these results, a discussion of the physical interpretation of the free data is presented in Section VII.  Finally Section VII concludes the presentation. 

\section{ADM 3+1 splitting of spacetime}
Beginning with Einstein's field equations that couple spacetime geometry to matter and non-gravitational fields
\begin{equation}\label{eqn:efe}G_{\mu\nu}=8 \pi T_{\mu\nu}\end{equation}
(here $G=c=1$, $G_{\mu \nu}$ is the Einstein tensor and $T_{\mu \nu}$ is the energy-momentum tensor), one can use a Cauchy IVP formulation with spatial hypersurfaces evolving along a congruence of timelike curves.  In the Arnowitt-Deser-Misner (ADM) framework \cite{alcubierre:3p1num}\cite{ADM}\cite{bernstein-phd}\cite{evans-phd}\cite{MTW} the dynamic variables that completely describe the spacetime are the spatial 3-metric $\gamma_{ab}$, gauge variables ($\alpha,\beta^i$) called the lapse function and shift vector respectively, and the extrinsic curvature $K_{ab}$.  The extrinsic curvature is defined in terms of a vector field $\xi^a$ of tangents to the time-like curves threading the spatial hypersurfaces by 
$$K_{ab} =  \nabla_a \xi_b.$$
In what follows Latin indices will be used for 3D spatial quantities and Greek indices for 4D space-time quantities.  The  gauge variable $\alpha$  measures the progression of proper time $\tau$ from one hypersurface to the next, i.e. $d\tau=\alpha dt$ at each point on the hypersurface, where $t$ is the coordinate time that can be used to label each spatial hypersurface.  The shift vector, $\beta^a$ measures the spatial shift of coordinates from one hypersurface to the next, and is a purely spatial quantity.

The full ADM 3+1 metric has the form
$$ds^2= - \alpha^2 dt^2 + \gamma_{ab}(dx^a+\beta^a dt)(dx^b + \beta^b dt)$$
The 3-spatial metric $\gamma_{ab}$ will be written in terms of spherical polar coordinates $(\eta, \theta, \varphi)$ and the line element takes the form:
\begin{equation}\label{eqn:3metricfinal}
dl^2=e^{4\phi}\left[e^qf_{,\eta}^2 d\eta^2 + e^qf^2d\theta^2 + f^2\sin^2{\theta}d{\varphi}^2  \right]
\end{equation}
where the functions $\phi$ and $q$ are functions of $\eta$ and $\theta$ and a subscript with a comma indicates partial differentiation with respect to the coordinate indicated in the subscript.
The radial variable $r=f(\eta)$ is chosen to allow for general distributions of radial positions for different numerical considerations.  For example setting $r=\sinh(\eta)$ places more numerical grid points near the origin in order to resolve the wave dynamics in the strongly non-linear regime. The choice of spherical polar coordinates instead of cylindrical coordinates is made for their natural ability to provide spherical radiative boundary conditions in the asymptotic region.  It should be noted that the metric used in equation (\ref{eqn:3metricfinal}) is not conformally flat, so numerical methods and slicings appropriate to conformally flat 3-metrics do not apply here.

The mixed form of the extrinsic curvature tensor can be written as
\begin{eqnarray}\label{eqn:mixkijtensor}
K^a_{\;b} & = & \left[\begin{array}{ccc}
H_a & H_c & 0 \\
\displaystyle{\frac{f_{,\eta}^2}{f^2}} H_c & H_b  & 0 \\
 0 & 0 & H_d \end{array} \right]
\end{eqnarray}
where all $H_a$, $H_b$, $H_c$, and $H_d$ are functions of $(\eta,\theta,t)$.

The Einstein field equations (\ref{eqn:efe}) lead to evolution equations for the extrinsic curvature variables, and the metric evolution equations follow as a geometric property of how the extrinsic curvature is defined. To complete the dynamical formulation a set of constraint equations called the Hamiltonian (scalar) and momentum (vector) constraints provide a means for determining a self-consistent set of initial values for the metric and extrinsic curvature.

The metric evolution equation has the general form
\begin{equation}\partial_t \gamma_{ab}=-2\alpha K_{ab}+D_a\beta_b+D_b\beta_a.
\label{eqn:3p1:gammadot}\end{equation}
where $D_a$ is the 3D covariant derivative operator with connection coefficients formed from $\gamma_{ab}$.  In the \emph{vacuum}, $T_{\mu \nu} = 0$. Then the general evolution equations for the mixed form of the extrinsic curvature are \cite{evans-phd}
\begin{eqnarray}\label{eqn:genmixkevol}
\partial_t K^a_{\;b} & = & -D^a D_b \alpha + \alpha [ R^a_{\;b} + K K^a_{\;b} ] \nonumber \\
& & + \beta^lD_l K^a_{\;b} + K^a_{\;l}D_b\beta^l - K^l_{\;b}D_l\beta^a. \label{eqn:3p1:Kevol3}\end{eqnarray}
where $K={\rm Tr}K=K^a_a$ and $R_{ab}$ is the Ricci tensor formed from the three-metric $\gamma_{ab}$.  After constructing the three-Ricci scalar $R= \gamma_{ab}R^{ab}$ 
the vacuum Hamiltonian constraint is given by
\begin{equation}R+({\rm Tr}K)^2-K^{ab}K_{ab}=0\label{eqn:3p1:ham3}\end{equation}
and the vacuum momentum constraints are given by
\begin{equation}D_b (K^{ab}-\gamma^{ab}{\rm Tr}K) = 0\label{eqn:3p1:mom3}\end{equation} 

\section{The Brill Wave Initial Conditions}
Brill's paper \cite{Brill} demonstrates that under the following conditions the mass-energy of the initial hypersurface is positive definite:
\begin{itemize}
\item The spacetime is axisymmetric ($\partial/\partial{\varphi}$ is a Killing vector)
\item The energy-momentum tensor vanishes ($T_{\mu\nu}=0$)
\item The initial space-like slice occurs at a moment of time symmetry for the metric ($\partial_t\gamma_{ab}=0$) such that $R=0$
\item The metric variable $q=q(r,{\theta})$ falls off faster than $1/r$ as $r\rightarrow \infty$
\item $q(r,0)=0$
\item The metric variables $q$ and $\phi$ are symmetric about the plane $z=0$ (${\theta}=\pm {\pi}/{2}$)
\end{itemize}

The simplest way to ensure that the time derivative of all metric quantities vanish on the initial slice is to set the shift vectors $\beta_i=0$ and have a vanishing extrinsic curvature $K_{ab}=0$ according to equation (\ref{eqn:3p1:gammadot}).  This also implies, via the Hamiltonian constraint 
(equation (\ref{eqn:3p1:ham3})) that 
$R=0$, which is consistent with the Brill formulation. 
As all the $K_{ab}=0$ initially, the momentum constraints (\ref{eqn:3p1:mom3}) are trivially satisfied.
%

%
In this formulation the initial value problem for Brill waves then reduces to a two-step process: (i) specifying a metric function $q$ subject to the conditions above, and (ii) solving the Hamiltonian constraint (6) for the metric variable $\phi$.

\subsection{Form of the metric function $q$}
There are an infinite variety of functions that will satisfy the Brill conditions above, however one additional requirement must be specified
to ensure regularity of the evolution equations: $q \sim r^n$ ($n\ge 2$) as $r \rightarrow 0$ (see \cite{MastersonPHD}).  Three general forms of $q$ were tested. Firstly a 
Gaussian form:
\begin{equation}\label{eqn:qinit}q(\eta,\theta,t=0)=2 A f^4e^{-\left({f}/{s_0}\right)^2}\sin^2\theta,\end{equation}
where the factor 2 multiplying the amplitude is introduced to provide for an easy comparison to ``traditional'' Brill wave formulations, $A$ is the ``amplitude'' of the wave and $s_0$ is a ``Gaussian RMS width'' parameter.  The second form of $q$ is an $\eta$ modulated version of the above form:
\begin{equation}\label{eqn:qinitsinmod}q(\eta,\theta,t=0)= \sin^2(4\eta)\left[2 A f^4e^{-\left({f}/{s_0}\right)^2}\sin^2\theta\right]\end{equation}
The third form is trigonometric in $\eta$ instead of exponential:
\begin{equation}\label{eqn:qinitsinpoly}q(\eta,\theta,t=0)=2 A \left(\frac{\sin^4(5\eta)}{(5\eta)^2}\right)\sin^2\theta\end{equation}

\section{Solving the Hamiltonian Constraint for $\phi$}
Employing the metric from equation (\ref{eqn:3metricfinal}) the 3-Ricci scalar can be computed. 
Then using the extrinsic curvature from equation (\ref{eqn:mixkijtensor}) the Hamiltonian constraint takes the form: 
\begin{eqnarray} & & \left(\frac{f}{f_{,\eta}}\right)^2\phi_{,\eta\eta} + \phi_{,\theta\theta} +  \frac{f}{f_{,\eta}}\left[1+\frac{\partial}{\partial{\eta}}\left(\frac{f}{f_{,\eta}}\right)\right] \phi_{,\eta} + \cot(\theta) \phi_{,\theta}  \nonumber \\ \mbox{} & &
+ \left(\frac{f}{f,_{\eta}}\right)^2 \phi_{,\eta}^2 + \phi_{,\theta}^2 \nonumber \\ \mbox{} & &
 = \frac{1}{4}e^q[f^2(H_b H_d + H_a H_d + H_a H_b)-f_{,\eta}^2H_c^2] e^{4\phi} \nonumber \\ \mbox{} & &
- \frac{1}{8}\left[q_{,\eta\eta}\left(\frac{f}{f_{,\eta}}\right)^2+q_{,\eta}\left(\frac{f}{f_{,\eta}}\right)\frac{\partial}{\partial{\eta}}\left(\frac{f}{f_{,\eta}}\right)+q_{,\theta\theta}\right] \label{eqn:hamconphi}
\end{eqnarray}
On the initial slice the extrinsic curvature components all vanish ($H_a=H_b=H_c=H_d=0$), and $q$ is given analytically by one of the forms above, leaving a quasi-linear second order inhomogeneous elliptic equation to be solved for $\phi$.

\subsection{Numerical Methods}
A detailed explanation of the numerical methods, boundary conditions, regularisation techniques, convergence tests, etc. used can be found in \cite{MastersonPHD}.
This section provides an overview of the methods used to solve equation (\ref{eqn:hamconphi}).

First a finite difference grid is set up in the coordinates $(0\le\eta\le\eta_{\rm{max}},0\le\theta\le{\pi}/{2},0\le t\le t_{\rm max})$. All differential operators are approximated using fourth order correct finite differences and phantom grid points are employed at the boundaries.  Boundary conditions are generally imposed via symmetry conditions at the axis, the equator and at $\eta=0$. At the outer boundary, $\eta = \eta_{\rm{max}}$, the function $\phi$ is assumed to consist of separable spherical functions of the form
\begin{equation}\label{eqn:NsepOB}  \phi = \left[\frac{c_1}{f} + \frac{c_{2}}{f^{2}} + \ldots\right]T(\theta). \end{equation}
For the IVP $c_1$, $c_2$, etc are constants.  During the evolution they they will be time dependent (i.e.~they will differ on different spatial slices).

Rearranging equation (\ref{eqn:hamconphi}) to place the linear (in $\phi$ and its derivatives) terms on the left-hand-side together with the vanishing of the extrinsic curvature terms one obtains:
\begin{eqnarray} & & \left(\frac{f}{f_{,\eta}}\right)^2\phi_{,\eta\eta} + \phi_{,\theta\theta} +  \frac{f}{f_{,\eta}}\left[1+\frac{\partial}{\partial{\eta}}\left(\frac{f}{f_{,\eta}}\right)\right] \phi_{,\eta} + \cot(\theta) \phi_{,\theta} \nonumber \\ \mbox{} & &
 =  - \left(\frac{f}{f_{,\eta}}\right)^2 \phi_{,\eta}^2 - \phi_{,\theta}^2 \nonumber \\ \mbox{} & &
- \frac{1}{8}\left[q_{,\eta\eta}\left(\frac{f}{f_{,\eta}}\right)^2+q_{,\eta}\left(\frac{f}{f_{,\eta}}\right)\frac{\partial}{\partial{\eta}}\left(\frac{f}{f_{,\eta}}\right)+q_{,\theta\theta}\right] \label{eqn:hamconphiarranged}
\end{eqnarray}
Due to the quasi-linear nature of equation (\ref{eqn:hamconphiarranged}) an iterative convergence scheme of the form 
$$U(\phi_{n+1}(i,j,t_m))=V(\phi_{n}(i,j,t_m))$$
is used
where the function $U$ is the linear elliptic operator that operates on $\phi$ and the function $V$ represents the non-linear operator operating on $\phi$. The
integer $n$ represents the looping/iteration counter used in the solver subroutine, $i$ is the radial discretisation index, $j$ the angular discretisation index and $t_m=0$ for the IVP.

A stabilized bi-conjugate gradient algorithm \cite{vandervorst}  was implemented to solve the matrix equation that results from a discretisation of equation (\ref{eqn:hamconphiarranged}). This technique  performs as expected across a wide range of scenarios.  Condition numbers are on the order of $K_1 \sim 10^6 \;;\; K_\infty \sim 10^5$ for the grids used ($200$ radial by $60$ angular divisions), and they increase as the number of grid points increase, indicating that this is an ill-posed numerical problem. Thus increasing the
number of grid points will not necessarily yield better results.  This has been analytically demonstrated when numerically solving Laplace's equation in certain situations (i.e. \cite{dijkstra-laplace}), so the result is consistent with theoretical findings.

In summary, the methodology employed for finding a complete IVP solution for Brill waves involves specifying a shape, amplitude and width for the initial $q$ function. A solution to the quasi-linear elliptic equation (\ref{eqn:hamconphiarranged}) can be obtained for the other metric variable $\phi$.  Once the solution is found, one can perform a search of the initial slice to determine whether or not the IVP contains a black hole. Intuitively one would expect that highly concentrated, large amplitude gravitational waves should cause a black hole to be present in the IVP.

\section{Finding Trapped Surfaces}
Given that the solution to the IVP is unable to to locate an event horizon due to its global-in-time property, the methodologies of \cite{cook-phd}\cite{bernstein-phd} that search for apparent horizons will be employed.  While it is possible that the subsequent evolution of our spacetime away from the IVP will reveal the presence of an event horizon, the methodology to be implemented here allows for local (in space and time) measures of the existence of black holes and the verification of theoretical predictions regarding black hole formation by gravitational waves.  It does not, however, allow one to say that a black hole is \emph{not} present; merely that an apparent horizon has not been detected.

To locate an apparent horizon one looks for a ``trapped'' 2-surface $S$ which is orientable and compact, and whose outward pointing spatial unit normal $s^a$ satisfies
\begin{equation}\label{eqn:trappedsurf}D_a s^a + K_{ab}s^a s^b - {\rm Tr}K = 0 \end{equation}
The outermost such \textbf{closed} trapped surface will form an apparent horizon that guarantees the creation of an event horizon within a finite time.
Assigning a measure of mass to the total area of the horizons that are formed one can expect that the ``mass'' of the event horizon is greater than or equal to  the ``mass'' of the apparent horizon.  This apparent horizon surface is also called a marginally outermost trapped surface (MOTS).

In general a surface $S$ can be parameterised on the hypersurface as
$$S=[\eta(l),\theta(l)],$$
where $l$ is a parameterisation variable. It will be assumed that the surface $S$ is topologically spherical, i.e.~it is not double valued at some interval
of the angular coordinate $\theta$ (which is verified numerically except in extreme cases). Therefore $S$ can determined by a radial function with only a
$\theta$ dependence in the less general format:
\begin{equation}\label{eqn:sparam}S=[h(\theta),\theta]\end{equation}
In this case equation (\ref{eqn:trappedsurf}) provides a means for solving for the function $h(\theta)$:
\begin{eqnarray}\label{eqn:trapsurffull}
 & & h_{,\theta\theta} + \left[\cot\theta+\frac{q_{,\theta}}{2}+4\phi_{,\theta}\right] h_{,\theta}   \nonumber \\ 
& &- \left[\frac{q_{,\eta}}{2} + 4\phi_{,\eta} + \left(\frac{f_{,\eta}}{f}\right)\left(2+\partial_\eta\left(\frac{f}{f_{,\eta}}\right)\right)\right] h_{,\theta}^2    \nonumber \\
& &+ \left(\frac{f_{,\eta}}{f}\right)^2\left[\cot\theta + \frac{q_{,\theta}}{2}+4\phi_{,\theta}\right] h_{,\theta}^3  \nonumber \\
& &+ \frac{f^2e^{2\phi-q}}{f_{,\eta}^4}[H_d+H_b+H_a] \delta^3  \nonumber \\
& &+ e^{2\phi} \left[-H_b h_{,\theta}^2+2H_ch_{,\theta}-\left(\frac{f}{f_{,\eta}}\right)^2H_a\right] \delta  \nonumber \\
&  & -\left(\frac{f}{f_{,\eta}}\right)^2\left[\frac{q_{,\eta}}{2}+4\phi_{,\eta}\right]-2\left(\frac{f}{f_{,\eta}}\right) = 0\;\;  
\end{eqnarray}
where $\delta$ is defined as
$$\delta=\delta(h(\theta))=f_{,\eta} e^{\frac{q}{2}}\sqrt{\left(\frac{f_{,\eta}}{f}\right)^2h_{,\theta}^2+1}$$
This is a quasi-linear equation for $h$ that is discretised using second order correct finite differences.  We then start at the equator and use the boundary condition $h_{,\theta}\left({\pi}/{2}\right)=0$ at one of our known grid points $h\left({\pi}/{2}\right)=\eta(i)$ to start iterating angularly for solutions (using a
standard ``shooting'' method), then repeat the method for each radial grid point in order to obtain a trapped surface map.  While some non-closed trapped surfaces present elsewhere in the grid might be missed with this methodology, the form of $q$ chosen will attain its maximum values on the equator and any fully enclosed surface will be found provided that the grid spacing is small enough.
\begin{figure}[h!] \begin{center}
{\includegraphics [scale = .70] {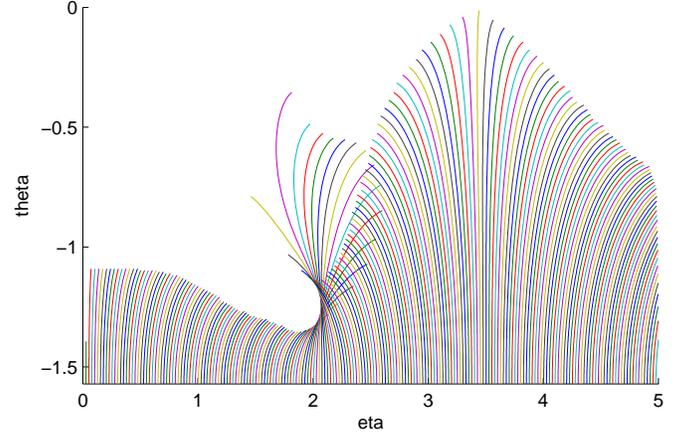}}
\caption{\label{fig:a9s01expIVP}Trapped surface structure for exponential IVP $(A,s_0,\eta_{\mathbf{max}})=(9,1,5)$ using equation (\ref{eqn:qinit})}\end{center} \end{figure}

\begin{figure}[h!] \begin{center}
{\includegraphics [scale = .65] {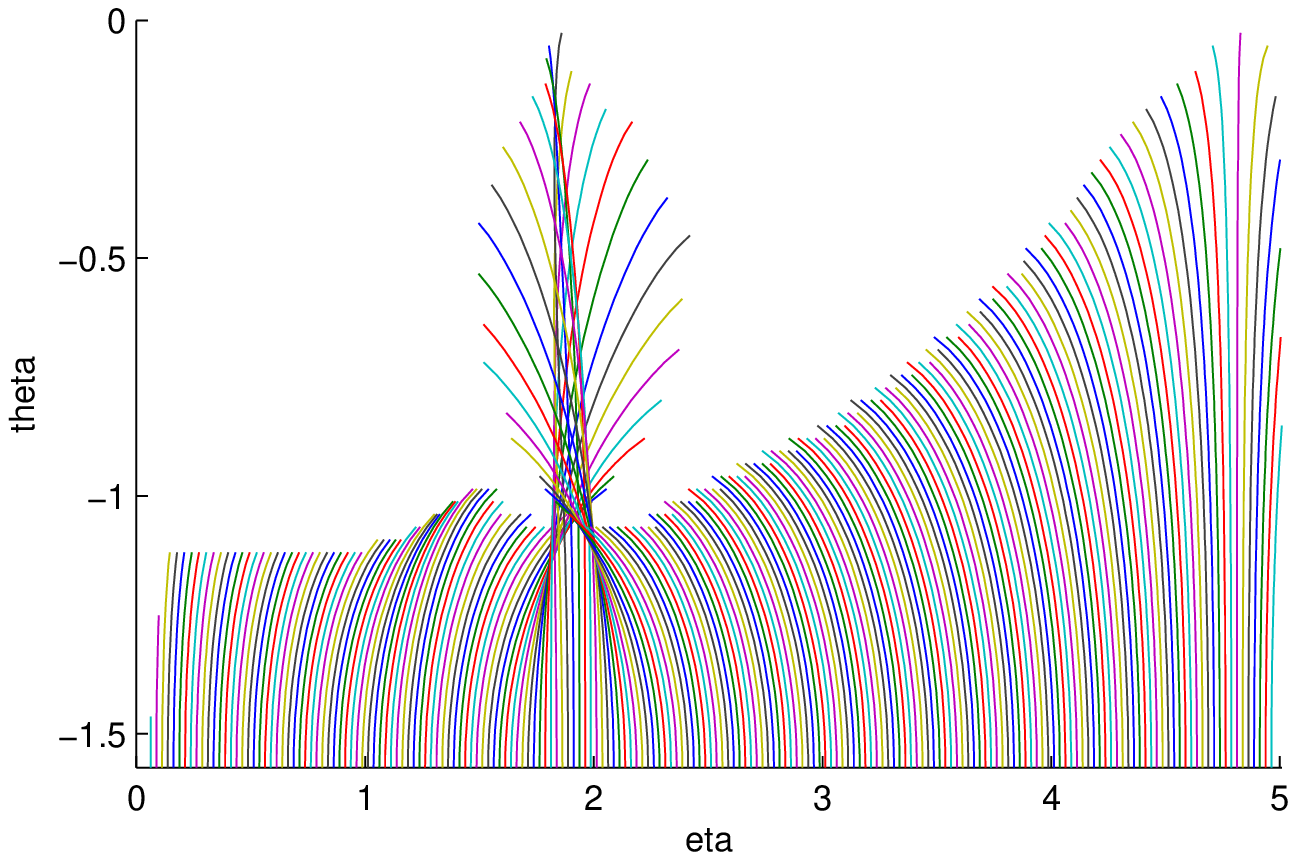}}
\caption{\label{fig:a-3.5s01modexpIVP}Trapped surface structure for modulated exponential IVP $(A,s_0,\eta_{\mathbf{max}})=(-3.5,1,5)$ using equation (\ref{eqn:qinitsinmod})}
\end{center} \end{figure}

\begin{figure} [h!]  \begin{center}
{\includegraphics [scale = .65] {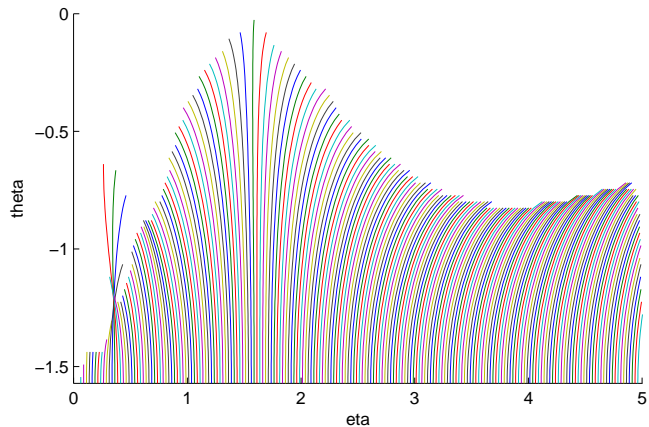}}
\caption{\label{fig:a10s01trigIVP}Trapped surface structure for trigonometric IVP $(A,\eta_{\mathbf{max}})=(10,5)$ using equation (\ref{eqn:qinitsinpoly})}
\end{center} \end{figure}

While others (e.g.~\cite{bernstein-phd},\cite{cook-phd}) attempt to solve this equation for a single closed trapped surface, we wish to study the trapped surface structure itself, for it yields important insights into the geometry of the spacetime being studied.

An example of the trapped surface structure that was found using the Gaussian form (\ref{eqn:qinit}) for $q$ with initial parameters 
$(A,s_0,\eta_{\mathbf{max}})=(9,1,5)$ (a very large amplitude wave) can be seen in Figure \ref{fig:a9s01expIVP}. 
 An example of the trapped surface for a $q$ function given by equation (\ref{eqn:qinitsinmod}) can be seen in Figure \ref{fig:a-3.5s01modexpIVP}, and similarly the trapped surface topology for $q$ using the trigonometric form in equation (\ref{eqn:qinitsinpoly}) can be seen in Figure \ref{fig:a10s01trigIVP}.
%
%
%
%
The nested curves on the plots give a map of partially formed trapped surfaces that are found by iterating angularly outwards from the equator at various radial points.  A fully formed closed trapped surface will be represented by a curve that extends from the bottom of the graph to the top (i.e.~$\theta=-\frac{\pi}{2} \rightarrow \theta=0$,) which by applying the
symmetry conditions will form a closed surface separating the spacetime into the interior and exterior regions of a black hole.

One very interesting feature of this trapped surface topology is that outside of the outermost closed trapped surface the solutions for $h$ will, for increasing $\theta$ 
tip outwards. As the outermost closed trapped surface (the apparent horizon) is approached, the curves will become more oriented along a constant value of $\eta$. 
Immediately inside the black hole the curves tip in the inward direction.  Deep inside the black hole it can be seen that there are regions where there is a 
transition between ``in-going'' and ``out-going'' regions separated by inner closed trapped surfaces.  This method of analyzing the trapped surface topology is useful
for exploring the interior of the black hole and in detecting when and where an apparent horizon is about to form in the numerical evolution of the system. 
\begin{figure} [h!] \begin{center}
{\includegraphics [scale = .55] {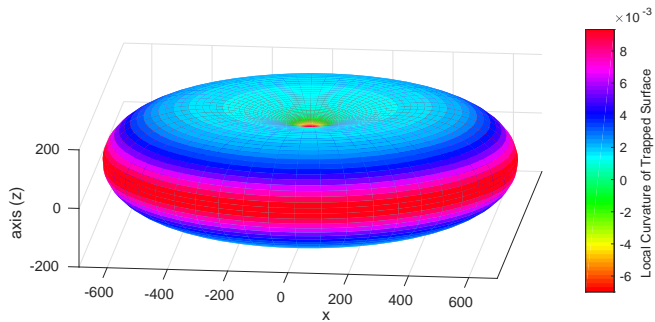}}
\caption{\label{fig:a9s01expIVPembed}Apparent horizon structure for exponential IVP $(A,s_0,\eta_{\mathbf{max}})=(9,1,5)$ using $\bar{r}$ embedding}\end{center}  \end{figure}

\begin{figure}[h!] \begin{center}
 {\includegraphics [scale = .90] {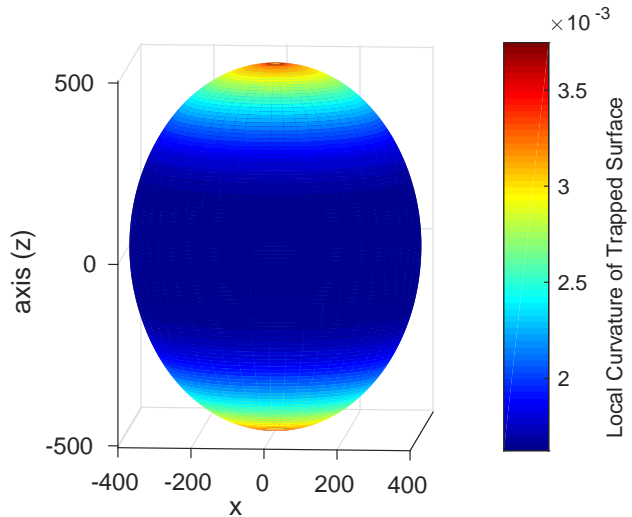}}
 \caption{\label{fig:a-4s01expIVPembed}Apparent horizon structure for exponential IVP $(A,s_0,\eta_{\mathbf{max}})=(-4,1,5)$ using $\bar{r}$ embedding}\end{center}  \end{figure}

\subsection{Embedding Trapped Surfaces}
Given a particular solution for the 3-metric it is useful to  determine how the trapped surfaces appear in a flat embedding manifold.  To this end, one can define a radial coordinate
\begin{equation}\label{eqn:spacetimerdist}\bar{r}=\int_{0}^{h(\theta)} \sqrt{\gamma_{11}} d\eta\end{equation}
which measures the radial spacetime distance from the origin to each point on a trapped surface.  This gives a sense of how a horizon is distorted or stretched 
anisotropically in a flat embedding manifold.

If the trapped surface is completely closed (i.e. the solver finds a trapped surface that goes from $\theta=0$ to $\theta={\pi}/{2}$) an apparent horizon
exists. Taking advantage of the axisymmetry and reflection symmetry across the equator the solution to equation (\ref{eqn:trapsurffull}) can be used
to construct an embedding diagram which produces a 3D visualization of the appearance of the apparent horizon. 
An example of this embedding for the traditional exponential data set $(A,s_0,\eta_{\mathbf{max}})=(9,1,5)$ can be seen in Figure \ref{fig:a9s01expIVPembed}.  An example for the negative amplitude wave $(A,s_0,\eta_{\mathbf{max}})=(-4,1,5)$ can be seen in Figure \ref{fig:a-4s01expIVPembed}.
%

In general it is found that large positive values of $A$ lead to oblate shaped embeddings with small dimples at the poles while large large negative values of 
the amplitude parameter lead to prolate ellipsoids. 

One can calculate an embedding of $\gamma_{11}$ on the horizon by setting $d\theta=0$ and equating the metric to a cylindrical flat-space metric that is a function of $(\rho,z)$.  This leads to the spatial portion of the metric being identified as
$$d\sigma^2 = \left[\left(\frac{dz}{d\eta}\right)^2 + \left(\frac{d\rho}{d\eta}\right)^2 \right]d\eta^2 + \rho^2d\varphi^2 = \gamma_{11} d\eta^2 + \gamma_{33}d\varphi^2$$
which yields
$$\rho=f e^{2\phi} \;;\; \frac{dz}{d\eta}=\sqrt{e^{q+4\phi}\left.f_{,\eta}\right.^2 - e^{2\phi}\left(2f\phi_{,\eta}+f_{,\eta}\right)}$$

\begin{figure}[h!] \begin{center}
{\includegraphics[scale = .68]{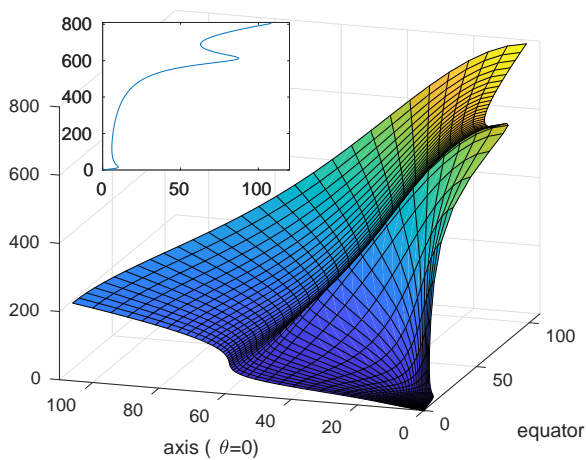}}
\caption{\label{fig:g11_embed_a9}Embedding diagram of $\gamma_{11}$ on the apparent horizon for various $\theta$=constant zones of an extreme positive amplitude $(A=9)$ Brill wave.  Inset plot shows the embedding at the equator $\left(\theta=\frac{\pi}{2}\right)$.}\end{center}  \end{figure}

\begin{figure}[h!] \begin{center}
{\includegraphics[scale = .68] {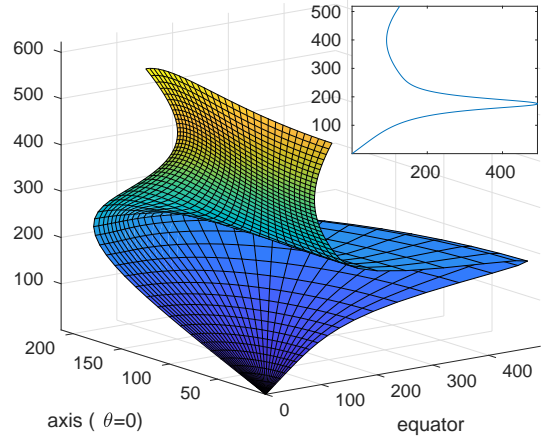}}
\caption{\label{fig:g11_embed_a-4}Embedding diagram of $\gamma_{11}$ on the apparent horizon for various $\theta$=constant zones of an extreme negative amplitude $(A=-4)$ Brill wave.  Inset plot shows the embedding at the equator $\left(\theta=\frac{\pi}{2}\right)$.}\end{center}  \end{figure}

These embeddings show that there are regions where the radial distance $\rho$ is a local minimum.  The outermost minimal radius leads to the formation of the outermost trapped
surface (apparent horizon). For the extreme amplitude $A=9$ this form of the embedding is shown in Figure \ref{fig:g11_embed_a9}. Inside the black hole there is a second minimal radius which is consistent with the horizon structure picture shown in Figure \ref{fig:a9s01expIVP}. It should be noted that as the amplitude $A$ is increased the inner minimal surface
becomes smaller and smaller. However before it collapses to zero (and therefore forming the so-called ``bag of gold'' singularity) \cite{wheeler-1964-lects} a new minimal surface forms outside of
of the first.  This means that a new apparent horizon forms at a finite radius.  A similar embedding diagram is shown for the amplitude $A=-4$  in Figure \ref{fig:g11_embed_a-4}.  This diagram is consistent
with the prolate spheroid shown in Figure \ref{fig:a-4s01expIVPembed}. The structure of the interior of the black
hole is such that there is a region where the proper radius can be much larger than that immediately outside of the apparent horizon.  In all cases no ``bag of gold''
singularity is observed forming at the outermost minimal surface.  This type of singularity does not form in the interior of any of the IVP black hole solutions.  

%
%
%
\section{Theoretical predictions}
A mathematical analysis of the dependence of the IVP solution on the amplitude parameter $A$ has been presented by {\'O} Murchadha in \cite{murchada-1993dgr2.conf..210O}.  He deduces that $A$ should have a maximum and minimum value ($A_+$ and $A_-$, respectively) for which the IVP has a solution, determined in \emph{general} by the shape of the specified $q$ function.  This arises due to Cantor and Brill's inequality \cite{cantor.brill}  which must be satisfied for a conformal mapping to exist (i.e. for $\phi$ to have a solution):
$$\int[8(\nabla u)^2+R(\bar{\gamma})u^2]dv>0$$
for every non-zero function $u$ with compact support. Here $\bar{\gamma}$ is the metric conformally related to $\gamma$ ($\gamma=e^{4\phi}\bar{\gamma}$) and $R(\gamma)=0$.  $R(\bar{\gamma})$ can be crafted (with a large enough positive or negative amplitude for $q$) to violate this inequality as it is the only term that is not strictly positive.

This means that there is a bounded parameter space in $A$ for which the Hamiltonian constraint (\ref{eqn:hamconphi}) has a solution given a specific profile for $q$. Outside those bounds one expects that the numerical solver for $\phi$ will be unable to converge as the apparent horizon that was present in the sub-critical IVP has now moved out to infinity and the spacetime then becomes undefined.

Furthermore it can be deduced, since $A=0$ is necessarily included in this parameter space (and flat space has no apparent horizon present), that there must be a region of small enough values of $A$ for which there is no apparent horizon present in the IVP. This would imply the existence of a critical value of
$A$ for which an apparent horizon first appears with almost fully trapped surfaces present in the sub-critical cases.  Indeed these phenomena are observed in our IVP solutions, providing a nice match with theory and a positive test of the numerical code.
\begin{figure*} \begin{center}
{\includegraphics[scale = .8] {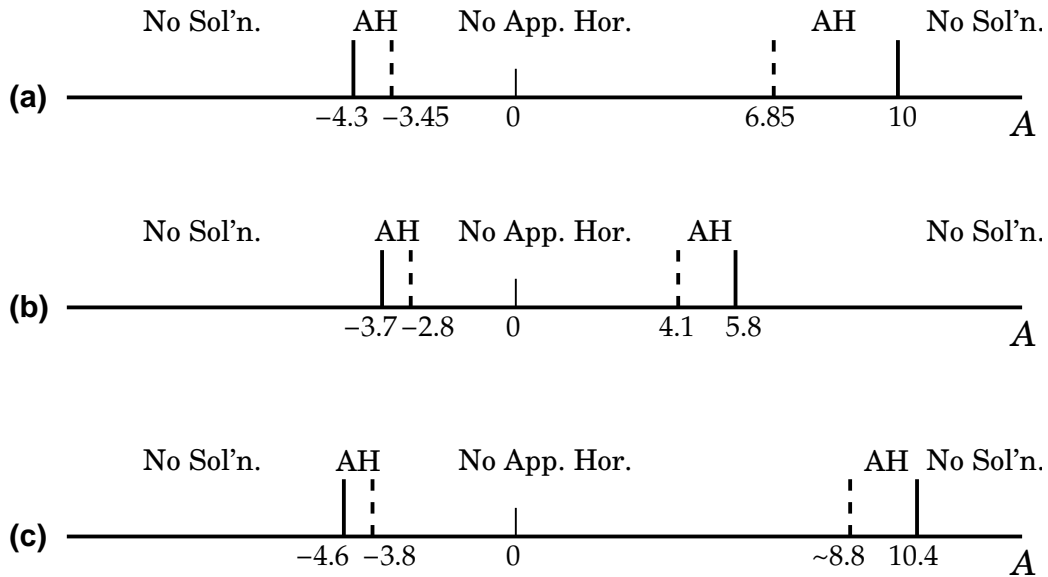}}
\caption{\label{fig:exp_phase_space} The parameter space mapping for the Brill wave amplitude $A$ as it appears in three different functional forms for the free data $q$. Equation (8) with $s_0=1$ leads to the figure (a), equation (9), also with $s_0 = 1$, to figure (b) and equation (10) to figure (c).  The mappings demonstrate that there is a small window of values for the amplitude of
the initial wave for which apparent horizons exist in solutions to the IVP. As the location of the apparent horizon moves outward toward infinity for the absolute value of both positive and negative values
of $A$, there is a critical value that determines whether or not a solution can be obtained for the initial value problem.}
\end{center} \end{figure*}

%
%
%

\subsection{Tested forms of $q$}
The numerical solver described above was tested on the three $q$ functions presented in equations (\ref{eqn:qinit}), (\ref{eqn:qinitsinmod}) and (\ref{eqn:qinitsinpoly}) to determine if we could observe the parameter space characterization as predicted by {\'O} Murchadha and described above.  More detailed figures and animations are presented in \cite{MastersonPHD}, and we achieved the predicted parameter space mapping for each of the functions. See Figure \ref{fig:exp_phase_space} %
for a depiction of the characterization of the changes in the amplitude $A$ and its affect on the solutions to the IVP.  As is typical of critical behaviour, as the
values of $A_-$ or $A_+$ are approached large variations in the observed horizon placement in $\eta$ coordinates (which lead to exponential exponential increases in the coordinate $r$) can be seen; i.e. a small change in the amplitude $A$ can cause a horizon to move several orders of magnitude further out in $r$ coordinates.

In all cases we observe that there is indeed a critical value of $A$ for which there is no numerical solution if $A<A_-$.  Similarly there is a maximum value $A_+$ beyond which the Hamiltonian constraint cannot be solved. Therefore we do not observe a periodicity in the parameter space as conjectured by \cite{murchada-1993dgr2.conf..210O}. This can be explained by the positive definiteness of the hypersurface mass-energy.

As the critical values in the parameter space are approached apparent horizon formation ``turns on''. Additionally the trapped surface topology ``tips up'' to eventually extending from $\theta=0$ to $\theta=\frac{\pi}{2}$.  Once an apparent horizon is present, increasing the value of $|A|$ further causes the horizon to migrate outwards in $\eta$ coordinates, eventually causing it to leave the numerical grid and move off to infinity.

\section{Physical Interpretation of $q$}
To understand the physical behaviour of changing the amplitude $A$ in the function $q$, and what positive and negative values of $A$ mean, let us return to the notion that the metric measures spacetime distances.  From equation (\ref{eqn:3metricfinal}), the vanishing of the shift vector on the initial hypersurface and the fact that $ds^2=0$ for light rays we note that the metric has the general form
\begin{equation}\label{eqn:metricqsimp}\alpha^2 dt^2 = e^{4\phi}\left[L e^q + P\right]\end{equation}
where $L=L(f,f_\eta,d\eta,d\theta)$ and $P=P(f,\theta,d{\varphi})$.
This means that $q$ is a cost function for relative motion of a light ray in the $\eta,\theta$ plane versus spinning around ${\varphi}$ (lines of latitude).

In areas with large \emph{positive} $q$ (where the apparent horizon has an oblate geometry), light rays will prefer motion along lines of latitude to minimize the contribution from $L$, which can be thought of as large relative curvature impedance in the $\eta$ and $\theta$ directions.

In areas of large \emph{negative} $q$ (i.e.~prolate apparent horizon geometry) light rays will prefer to move in the $\eta,\theta$ plane to minimize the contribution of $P$, and therefore avoid rotation about the axis of symmetry.  This also means that there is large relative curvature impedance in the ${\varphi}$ direction.

After we determine the relative cost of motion in these directions by specifying $q$, the Hamiltonian constraint is then solved for $\phi$ which provides the correct global scaling factor so that equation (\ref{eqn:metricqsimp}) holds.

That there is a limit on the amplitude of $q$ (as discussed in the context of {\'O} Murchadha's paper above) for which a solution to $\phi$ exists, indicates 
that black holes formed from a concentration of Brill waves are only so ``strong'' - i.e. there is a critical relative curvature impedance between the $\eta,\theta$ plane and ${\varphi}$ that cannot be exceeded even with a black hole of infinite extent.

\section{Conclusion}
We have discussed a numerical framework for constructing solutions to the Brill wave IVP, with a variety of shapes and amplitudes for the free specified metric function.  The solutions that were obtained have been compared to theoretical work by {\'O} Murchadha and found to be in agreement, as well as providing an insight into the topology of trapped surfaces in the spacetime.  A critical limit on the value of the amplitude appearing in the function $q$ of a Brill spacetime is also discussed, indicating that black holes have limits to their strengths.

\section{Acknowledgements}
This work as sponsored in part by a Discovery Grant from the Natural Sciences and Engineering Research Council of Canada.  The authors would like to thank, Miguel Alcubierre,
Matt Choptuik, David Garfinkle and Franz Praetorius for discussions regarding methods in numerical relativity and in particular those applied to Brill waves that have influenced approaches 
taken in this work.

\bibliography{mybib4}{}
\bibliographystyle{plain}
\end{document}